\documentclass{article}
\usepackage{cite}
\usepackage{graphicx}
\usepackage{dcolumn}

\begin{document}

\date{}
\title{Variational approach to the Schr\"{o}dinger equation with a delta-function
potential}
\author{Francisco M. Fern\'{a}ndez\thanks{fernande@quimica.unlp.edu.ar} \\
INIFTA, DQT, Sucursal 4, C.C 16, \\
1900 La Plata, Argentina}

\maketitle

\begin{abstract}
We obtain accurate eigenvalues of the one-dimensional Schr\"{o}dinger
equation with a Hamiltonian of the form $H_{g}=H+g\delta (x)$, where $\delta
(x)$ is the Dirac delta function. We show that the well known Rayleigh-Ritz
variational method is a suitable approach provided that the basis set takes
into account the effect of the Dirac delta on the wavefunction.
\end{abstract}

\section{Introduction}

\label{sec:intro}

A quantum-mechanical Hamiltonian operator $H$ perturbed by a delta-function
potential $g\delta (x)$ has received considerable attention\cite
{AC75,L82,R85,L87,APST87,FI94,FI96,BERW98,P06,VP11,FB17,CLW18,GS20}. In most
cases $H$ describes a free particle\cite{AC75}, a particle in a box\cite
{AC75,L82,R85,L87} or the harmonic oscillator\cite
{AC75,APST87,FI94,FI96,BERW98,P06,VP11,FB17,CLW18,GS20}. Since in these
cases the Schr\"{o}dinger equation for $H$ is exactly solvable one can
obtain closed form expressions for the solutions to the Schr\"{o}dinger
equation for $H_{g}=H+g\delta (x)$ in several different ways. For example,
from the eigenvalues and eigenfunctions of $H$\cite{AC75,BERW98}, by solving
the eigenvalue equation left and right of the origin and matching those
solutions at $x=0$\cite{L82, R85,APST87,P06,VP11} or by means of the Green
function\cite{FB17,CLW18}. In some cases the authors resorted to this kind
of models to illustrate the application of approximate methods like
perturbation theory\cite{L87,P06}, WKM method\cite{P06}, or variational
approaches\cite{P06,GS20}. Several such proposals have proved of pedagogical
interest\cite{AC75,R85,L82,L87,P06,VP11,GS20} and a student may inquire
about the possibility of solving examples in which the Schr\"{o}dinger
equation for $H$ is not exactly solvable.The purpose of this paper is to
address this point.

In most undergraduate courses on quantum mechanics and quantum chemistry the
students become familiar with approximate methods like perturbation theory
or variational techniques. For example, the Rayleigh-Ritz variational method
is particularly useful in atomic and molecular physics\cite{SO96}. Here, we
show how to choose a suitable basis set that takes into account the effect
of the Dirac-delta-function potential. In section~\ref{sec:model} we outline
the model and some of the properties of the Schr\"{o}dinger equation. In
section~\ref{sec:Rayleigh-Ritz} we illustrate the application of the
Rayleigh-Ritz variational method to a family of polynomial potentials.
Finally, is section~\ref{sec:conclusions} we summarize the main results and
draw conclusions.

\section{The model}

\label{sec:model}

In what follows we restrict ourselves to the dimensionless Schr\"{o}dinger
equation\cite{F20}
\begin{equation}
\psi ^{\prime \prime }(x)=2\left[ V(x)+g\delta (x)-E\right] \psi (x),
\label{eq:Schro}
\end{equation}
where $\delta (x)$ is the Dirac delta function. The delta-function potential
determines the well known behaviour of the wavefunction at origin
\begin{equation}
\psi \left( 0^{-}\right) =\psi \left( 0^{+}\right) =\psi (0),\;\psi ^{\prime
}\left( 0^{+}\right) -\psi ^{\prime }\left( 0^{-}\right) =2g\psi (0).
\label{eq:cond_delta}
\end{equation}
According to the Hellmann-Feynman theorem\cite{G32,F39} every energy
eigenvalue increases with the strength parameter of the delta potential as
\begin{equation}
\frac{\partial E}{\partial g}=\left| \psi (0)\right| ^{2},\;\left\langle
\psi \right. \left| \psi \right\rangle =1.  \label{eq:HFT}
\end{equation}

If the potential-energy function is parity invariant ($V(-x)=V(x)$) then the
eigenfunctions are either even ($\psi _{e}(-x)=\psi _{e}(x)$) or odd ($\psi
_{o}(-x)=-\psi _{o}(x)$) and the Hellmann-Feynman theorem tells us that the
energies of the latter states do not change with $g$ because $\psi _{o}(0)=0$%
. In other words, the odd states are solutions to equation (\ref{eq:Schro})
with $g=0$ which is consistent with the fact that $\psi _{o}^{\prime }(x)$
is continuous at origin according to equation (\ref{eq:cond_delta}). The
behaviour of the even states at origin becomes
\begin{equation}
\psi ^{\prime }(0^{+})=g\psi (0),  \label{eq:cond_delta_even}
\end{equation}
and throughout this paper we consider that the wavefunction also satisfies $%
\psi (x\rightarrow \pm \infty )=0$.

When $|g|$ is sufficient small we can apply perturbation theory and obtain
an expansion of the form
\begin{equation}
E_{n}(g)=\sum_{j=0}^{\infty }E_{n}^{(j)}g^{j},  \label{eq:PT_g_small}
\end{equation}
where the coefficients $E_{n}^{(j)}$ can be obtained in closed form provided
that the eigenvalue equation for $H_{0}=H$ is exactly solvable. Examples are
given by the particle in a box\cite{L87} and the harmonic oscillator\cite
{P06}.

On the other hand, when $|g|\rightarrow \infty $ equation (\ref
{eq:cond_delta_even}) yields $\psi (0)=0$ and the solutions are the
odd-parity states of $H$. An exception is the ground state when $%
g\rightarrow -\infty $\cite{P06,VP11}. We will discuss this point with more
detail in the examples studied in section~\ref{sec:Rayleigh-Ritz}. Here, we
just mention that there is a critical value $g_{0}$ such that $E_{0}(g)>0$
if $g>g_{0}$ and $E_{0}(g)<0$ if $g<g_{0}$.

\section{The Rayleigh-Ritz approach}

\label{sec:Rayleigh-Ritz}

In order to apply the Rayleigh-Ritz variational method to the
Schr\"{o}dinger equation $H\psi =E\psi $ we choose a suitable basis set $%
\left\{ \varphi _{j},\;j=0,1,\ldots \right\} $ and construct the trial
function
\begin{equation}
\varphi =\sum_{j=0}^{N}c_{j}\varphi _{j}.  \label{eq:trial_funct}
\end{equation}
Then we obtain the minimum of the variational integral
\begin{equation}
W=\frac{\left\langle \varphi \right| H\left| \varphi \right\rangle }{%
\left\langle \varphi \right. \left| \varphi \right\rangle },
\label{eq:W_var}
\end{equation}
with respect to the expansion coefficients $c_{j}$%
\begin{equation}
\frac{\partial W}{\partial c_{j}}=0,\;j=0,1,\ldots ,N.  \label{eq:W_minima}
\end{equation}
This approach is well described in many textbooks\cite{SO96} so that we will
only show the results here. The expansion coefficients $c_{j}$ are solutions
to the \textit{secular equation }
\begin{eqnarray}
\sum_{j=0}^{N}\left( H_{ij}-WS_{ij}\right) c_{j} &=&0,\;i=0,1,\ldots ,N,
\nonumber \\
H_{ij} &=&\left\langle \varphi _{i}\right| H\left| \varphi _{j}\right\rangle
,\;S_{ij}=\left\langle \varphi _{i}\right. \left| \varphi _{j}\right\rangle ,
\label{eq:secular_eq}
\end{eqnarray}
and there are nontrivial solutions only for those values of $W$ that are
roots of the \textit{secular determinant}
\begin{equation}
\left| \mathbf{H}-W\mathbf{S}\right| =0,  \label{eq:secular_det}
\end{equation}
where $\mathbf{H}$ and $\mathbf{S}$ are $(N+1)\times (N+1)$ matrices with
elements $H_{ij}$ and $S_{ij}$, respectively.These roots $W_{j}^{[N]}$, $%
j=0,1,\ldots ,N$, are real and satisfy $W_{j}^{[N]}\geq W_{j}^{[N+1]}\geq
E_{j}$, where $E_{j}$ is an eigenvalue of $H$\cite{SO96}.

A suitable basis set for the class of polynomial potentials $V(x)$ discussed
here is given by the Gaussian functions
\begin{equation}
\varphi _{j}=x^{j}\exp \left( -\frac{ax^{2}}{2}\right) ,\;j=0,1,\ldots
,\;a>0.  \label{eq:Gaussian_basis}
\end{equation}
However, if we require the trial function to satisfy (\ref
{eq:cond_delta_even}) at origin then a more convenient basis set is
\begin{equation}
u_{1}(x)=\left( 1+gx\right) \exp \left( -\frac{ax^{2}}{2}\right)
,\;u_{j}=x^{j}\exp \left( -\frac{ax^{2}}{2}\right) ,\;j=2,3,\ldots ,\;x>0,
\label{eq:basis_set_delta}
\end{equation}
and the trial function now reads
\begin{equation}
\varphi (x)=\sum_{j=1}^{N}c_{j}u_{j}(x),\;x>0.  \label{eq:trial_funct_delta}
\end{equation}

For the application of the Rayleigh-Ritz variational method outlined above
to present models we resort to the scalar product
\begin{equation}
\left\langle F\right. \left| G\right\rangle =\int_{0}^{\infty
}F(x)^{*}G(x)\,dx,  \label{eq:scalar_prod_delta}
\end{equation}
because it is only necessary to take into account half the coordinate space,
for example, $0\leq x<\infty $, when $V(x)$ is parity invariant. For
simplicity, we restrict ourselves to monomial potentials of the form
\begin{equation}
V(x)=A|x|^{b},\;A,b>0,  \label{eq:V(x)_monomial}
\end{equation}
so that all the integrals appearing in $\mathbf{H}$ and $\mathbf{S}$ are of
the form
\begin{equation}
\int_{0}^{\infty }x^{s}\exp \left( -\frac{ax^{2}}{2}\right)
\,dx=2^{(s-1)/2}a^{-(s+1)/2}\Gamma \left( \frac{s+1}{2}\right) ,
\label{eq:Gaussian_integral}
\end{equation}
where $\Gamma (z)$ is the gamma function.

In order to test the approach we first choose the harmonic oscillator
\begin{equation}
V(x)=\frac{1}{2}x^{2},  \label{eq:V(x)_HO}
\end{equation}
because there are simple transcendental equations for its eigenvalues\cite
{AC75,APST87,FI94,P06,VP11,FB17}. Tables \ref{tab:RRHO_1} and \ref
{tab:RRHO_-1} show that the Rayleigh-Ritz variational results converge from
above to the exact eigenvalues when $g=\pm 1$. For simplicity, we have
chosen $a=1$ because it yields the correct asymptotic behaviour of $\varphi
(x)$ at infinity. The results of these tables strongly suggest that present
approach is sound, at least for moderate values of $|g|$.

Tables \ref{tab:RRAO_1} and \ref{tab:RRAO_-1} show results for the
anharmonic oscillator with $V(x)=x^{4}$ and $g=\pm 1$. Since $x^{4}\gg x^{2}$
for $x\gg 1$ we expect the eigenfunctions of the anharmonic oscillator to
vanish asymptotically more rapidly; consequently, in this case we
arbitrarily chose $a=2$ to take into account this fact. It would be better
to obtain the optimal value of $a$ variationally but it would make the
calculation rather more involved. Although in this case we do not have exact
results for comparison, we are confident about the accuracy of the results
because the roots of the secular equation clearly converge to a limit from
above.

As a final example we choose $V(x)=|x|^{3}$. Tables \ref{tab:RRAO3_1} and
\ref{tab:RRAO3_-1} show that the convergence of the variational results for $%
g=\pm 1$ is satisfactory when $a=2$. The roots of the secular determinant
converge from above and the accuracy of the results is expected to be of the
order of the last stable digit.

The Rayleigh-Ritz variational method yields accurate results also for large $%
|g|$, the only exception being the ground state when $g\rightarrow -\infty $%
. We can estimate this energy eigenvalue by means of perturbation theory if
we choose $V(x)$ to be the perturbation. In fact, by means of a simple
scaling argument\cite{F20} we can easily prove that
\begin{equation}
E_{0}(g)=-\frac{|g|^{2}}{2}+\Gamma (b+1)A|g|^{-b}+\sum_{j=2}^{\infty
}e_{j}A^{j}|g|^{-(b+2)j+2},  \label{eq:PT_g_large}
\end{equation}
for the family of potentials in equation (\ref{eq:V(x)_monomial}).

\begin{figure}[tbp]
\begin{center}
\includegraphics[width=9cm]{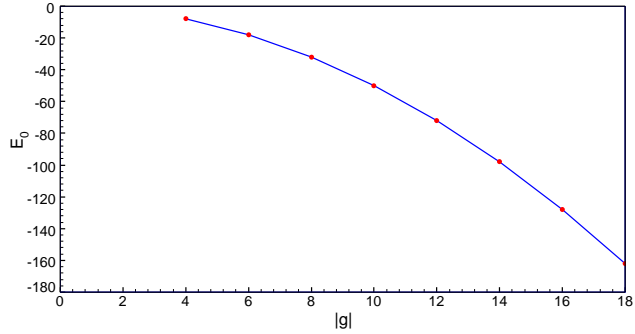}
\end{center}
\caption{$E_0(g)$ for the harmonic oscillator calculated by means of
perturbation theory (blue continuous line) and the Rayleigh-Ritz variational
method (red points) }
\label{Fig:E0x2PT}
\end{figure}

Figure~\ref{Fig:E0x2PT} shows that present Rayleigh-Ritz variational results
(with $N=17$) agree with the perturbation expression
\begin{equation}
E_{0}(g)\approx -\frac{g^{2}}{2}+\frac{1}{g^{2}},  \label{eq:E_0_PT_x2}
\end{equation}
for the harmonic oscillator (\ref{eq:V(x)_HO}) for moderately large values
of $|g|$. Larger values of $|g|$ will require larger values of $N$ in order
to obtain results of similar accuracy.

Finally, it is worth mentioning that the Rayleigh-Ritz variational method is
suitable for the calculation of the critical values $g_{0}$ mentioned at the
end of section~\ref{sec:model}. We simply set $W=0$ and solve the secular
determinant (\ref{eq:secular_det}) for $g$. We thus obtain $%
g_{0}^{H}=-0.6759782401$, $g_{0}^{Q}=-0.7515940253$ and $%
g_{0}^{C}=-0.7651281365$ for the harmonic, quadratic and cubic potentials,
respectively. The result for the harmonic oscillator agrees with the one
predicted by the exact analytical expression for the eigenvalues\cite
{AC75,BERW98}.

\section{Conclusions}

\label{sec:conclusions}

The results of this paper clearly show that the Rayleigh-Ritz variational
method is a suitable tool for the treatment of the Schr\"{o}dinger equation
perturbed by a Dirac-delta-function potential provided that the trial
function exhibits the correct behaviour at origin (or, in general, at the
location of the delta function). This behaviour can be easily introduced in
the basis set, at least for even-parity potentials. We have illustrated the
application of the approach by means of three monomial potentials and a
similar calculation for polynomial potentials is straightforward. The basis
set chosen is suitable for moderate values of $|g|$ as suggested by the
remarkable rate of convergence shown in tables \ref{tab:RRHO_1}-\ref
{tab:RRAO3_-1}. For large, positive values of $g$ the performance of the
variational method is similar because $\psi (0)\rightarrow 0$ as $%
g\rightarrow \infty $. The only difficulty may be found for the ground state
when $g\ll -1$ because the wavefunction is expected to behave asymptotically
as $\psi (x)\sim \exp \left( -g|x|\right) $. In this case it is required a
large basis set of Gaussian functions or a more convenient set of functions.
However, for most purposes the approach proposed here is sound.

The variational method proposed by Patil\cite{P06} and improved by Ghose and
Sen\cite{GS20} can also be applied to the models discussed above in the
preceding section. However, this approach, based on just one trial function
with adjustable parameters, only applies to the ground state. On the other
hand, the Rayleigh-Ritz method outlined in this paper yields estimates for
all the eigenvalues with the advantage that we can monitor the accuracy of
the results because the roots of the secular determinant converge to the
actual eigenvalues from above.

The application of the Rayleigh-Ritz variational method to problems of
physical interest commonly requires resorting to suitable computer software
for the calculation of the approximate eigenvalues and eigenfunctions. In
our opinion, this is a good opportunity for introducing the students to any
of the available computer-algebra software that enable one to calculate the
integrals in the matrix elements and provide algorithms for the solution of
the secular equations.

\begin{table}[tbp]
\caption{Eigenvalues for $V(x)=x^2/2$ and $g=1$}
\label{tab:RRHO_1}
\begin{center}
\par
\begin{tabular}{|rD{.}{.}{10}D{.}{.}{9}D{.}{.}{9}D{.}{.}{9}D{.}{.}{9}|}

\hline

$N$ &  \multicolumn{1}{c}{$W_0$}& \multicolumn{1}{c}{$W_1$} &
\multicolumn{1}{c}{$W_2$}&\multicolumn{1}{c}{$W_3$}&\multicolumn{1}{c}{$W_4$}\vline\\
\hline

 2 & 0.8934625502 &   2.760032981 &               &               &               \\
 3 & 0.8927956113 &   2.754728061 &   4.892590011 &               &               \\
 4 & 0.8927481911 &   2.754644627 &   4.700407469 &   7.555272621 &               \\
 5 & 0.8927444033 &   2.754641685 &   4.700197637 &   6.696400420 &   11.04476365 \\
 6 & 0.8927440777 &   2.754641542 &   4.700195867 &   6.669920689 &   8.835147079 \\
 7 & 0.8927440483 &   2.754641534 &   4.700195827 &   6.669909113 &   8.653953092 \\
 8 & 0.8927440456 &   2.754641533 &   4.700195826 &   6.669909053 &   8.650087699 \\
 9 & 0.8927440453 &   2.754641533 &   4.700195826 &   6.669909052 &   8.650086945 \\
10 & 0.8927440453 &   2.754641533 &   4.700195826 &   6.669909052 &   8.650086942 \\
11 & 0.8927440453 &   2.754641533 &   4.700195826 &   6.669909052 &   8.650086942 \\
\hline
\multicolumn{1}{c}{Exact}  &0.8927440453&2.754641533&4.700195826&6.669909052&8.650086942\\
\hline
\end{tabular}
\end{center}
\end{table}

\begin{table}[tbp]
\caption{Eigenvalues for $V(x)=x^2/2$ and $g=-1$}
\label{tab:RRHO_-1}
\begin{center}
\par
\begin{tabular}{|rD{.}{.}{10}D{.}{.}{9}D{.}{.}{9}D{.}{.}{9}D{.}{.}{8}|}

\hline

$N$ &  \multicolumn{1}{c}{$W_0$}& \multicolumn{1}{c}{$W_1$} &
\multicolumn{1}{c}{$W_2$}&\multicolumn{1}{c}{$W_3$}&\multicolumn{1}{c}{$W_4$}\vline\\
\hline

 2 &  -0.3085452475 &  2.222794094 &              &              &              \\
 3 &  -0.3397525400 &  2.220830361 &  4.340972321 &              &              \\
 4 &  -0.3421635872 &  2.220772352 &  4.291279595 &  6.777044277 &              \\
 5 &  -0.3423925293 &  2.220769679 &  4.291227875 &  6.332552744 &  9.910216406 \\
 6 &  -0.3424161154 &  2.220769524 &  4.291227060 &  6.325780145 &  8.440061646 \\
 7 &  -0.3424186377 &  2.220769513 &  4.291227036 &  6.325777482 &  8.348294919 \\
 8 &  -0.3424189127 &  2.220769513 &  4.291227035 &  6.325777458 &  8.347325939 \\
 9 &  -0.3424189430 &  2.220769513 &  4.291227035 &  6.325777457 &  8.347325766 \\
10 &  -0.3424189464 &  2.220769513 &  4.291227035 &  6.325777457 &  8.347325765 \\
11 &  -0.3424189467 &  2.220769513 &  4.291227035 &  6.325777457 &  8.347325765 \\

\hline
\multicolumn{1}{c}{Exact}  &-0.3424189467&2.220769513&4.291227035&6.325777457&8.347325765\\
\hline
\end{tabular}
\end{center}
\end{table}

\begin{table}[tbp]
\caption{Eigenvalues for $V(x)=x^4$ and $g=1$}
\label{tab:RRAO_1}
\begin{center}
\par
\begin{tabular}{|rD{.}{.}{9}D{.}{.}{9}D{.}{.}{8}D{.}{.}{8}D{.}{.}{8}|}
\hline $N$ &  \multicolumn{1}{c}{$W_0$}& \multicolumn{1}{c}{$W_1$}
&
\multicolumn{1}{c}{$W_2$}&\multicolumn{1}{c}{$W_3$}&\multicolumn{1}{c}{$W_4$}\vline\\
\hline

 2  & 1.212885159  & 5.431056449 &               &             &              \\
 3  & 1.204415202  & 5.289165764 &  11.68242949  &             &              \\
 4  & 1.203002874  & 5.189408545 &  11.21956836  & 19.89298470 &              \\
 5  & 1.202268166  & 5.188704722 &  10.75280031  & 18.84002195 &  30.32575650 \\
 6  & 1.202214950  & 5.180474979 &  10.70378223  & 17.54865753 &  28.38504206 \\
 7  & 1.202194697  & 5.180471582 &  10.66581963  & 17.25408723 &  25.68777289 \\
 8  & 1.202189533  & 5.179885928 &  10.66174091  & 17.14754048 &  24.72896521 \\
 9  & 1.202188331  & 5.179885715 &  10.65833898  & 17.09896256 &  24.52894335 \\
10  & 1.202188024  & 5.179845929 &  10.65788573  & 17.09042900 &  24.28708252 \\
11  & 1.202187948  & 5.179845863 &  10.65762856  & 17.08370796 &  24.27798766 \\
12  & 1.202187929  & 5.179842983 &  10.65757489  & 17.08321447 &  24.23689527 \\
13  & 1.202187923  & 5.179842956 &  10.65755810  & 17.08237213 &  24.23688451 \\
14  & 1.202187922  & 5.179842748 &  10.65755176  & 17.08235926 &  24.23152178 \\
15  & 1.202187922  & 5.179842741 &  10.65755086  & 17.08226559 &  24.23140156 \\
16  & 1.202187922  & 5.179842726 &  10.65755014  & 17.08226555 &  24.23085274 \\
17  & 1.202187922  & 5.179842725 &  10.65755010  & 17.08225643 &  24.23079034 \\

\hline
\end{tabular}
\end{center}
\end{table}

\begin{table}[tbp]
\caption{Eigenvalues for $V(x)=x^4$ and $g=-1$}
\label{tab:RRAO_-1}
\begin{center}
\par
\begin{tabular}{|rD{.}{.}{10}D{.}{.}{9}D{.}{.}{9}D{.}{.}{8}D{.}{.}{8}|}
\hline $N$ &  \multicolumn{1}{c}{$W_0$}& \multicolumn{1}{c}{$W_1$}
&
\multicolumn{1}{c}{$W_2$}&\multicolumn{1}{c}{$W_3$}&\multicolumn{1}{c}{$W_4$}\vline\\
\hline

 2  & -0.2444852177  & 4.409477284 &               &             &              \\
 3  & -0.2912647514  & 4.227287815 &  10.85766906  &             &              \\
 4  & -0.2928240157  & 4.182776600 &  10.15066069  & 19.13616079 &              \\
 5  & -0.2928292406  & 4.181284514 &  9.901826018  & 17.56949565 &  29.58232215 \\
 6  & -0.2928581504  & 4.176818119 &  9.846560349  & 16.77721127 &  26.77001070 \\
 7  & -0.2928677179  & 4.176745715 &  9.825535133  & 16.44942678 &  24.96731076 \\
 8  & -0.2928677613  & 4.176468061 &  9.821168085  & 16.39206046 &  23.91364185 \\
 9  & -0.2928678921  & 4.176464550 &  9.819348618  & 16.34614316 &  23.81216354 \\
10  & -0.2928679190  & 4.176446073 &  9.818909872  & 16.34209974 &  23.58709655 \\
11  & -0.2928679342  & 4.176445745 &  9.818783474  & 16.33634692 &  23.58472920 \\
12  & -0.2928679349  & 4.176444476 &  9.818737897  & 16.33616880 &  23.55051960 \\
13  & -0.2928679358  & 4.176444430 &  9.818730435  & 16.33550881 &  23.55040892 \\
14  & -0.2928679358  & 4.176444345 &  9.818725612  & 16.33550740 &  23.54632955 \\
15  & -0.2928679359  & 4.176444338 &  9.818725278  & 16.33543981 &  23.54613343 \\
16  & -0.2928679359  & 4.176444332 &  9.818724777  & 16.33543925 &  23.54575290 \\
17  & -0.2928679359  & 4.176444331 &  9.818724770  & 16.33543318 &  23.54568657 \\

\hline
\end{tabular}
\end{center}
\end{table}

\begin{table}[tbp]
\caption{Eigenvalues for $V(x)=|x|^3$ and $g=1$}
\label{tab:RRAO3_1}
\begin{center}
\par
\begin{tabular}{|rD{.}{.}{9}D{.}{.}{9}D{.}{.}{9}D{.}{.}{8}D{.}{.}{8}|}
\hline $N$ &  \multicolumn{1}{c}{$W_0$}& \multicolumn{1}{c}{$W_1$}
&
\multicolumn{1}{c}{$W_2$}&\multicolumn{1}{c}{$W_3$}&\multicolumn{1}{c}{$W_4$}\vline\\
\hline

 2 &  1.193494101 &  4.646154382 &               &             &              \\
 3 &  1.193483838 &  4.624404432 &  9.146712426  &             &              \\
 4 &  1.192694864 &  4.615701763 &  8.900591260  & 14.75342461 &              \\
 5 &  1.192687849 &  4.612065439 &  8.847299481  & 13.67700784 &  21.87335551 \\
 6 &  1.192686970 &  4.612046916 &  8.829421034  & 13.57433311 &  18.83682980 \\
 7 &  1.192686545 &  4.611976831 &  8.829268044  & 13.49526184 &  18.74190242 \\
 8 &  1.192686488 &  4.611976382 &  8.828629838  & 13.49466862 &  18.48745373 \\
 9 &  1.192686486 &  4.611975486 &  8.828628402  & 13.49109442 &  18.48701306 \\
10 &  1.192686486 &  4.611975428 &  8.828615220  & 13.49108590 &  18.47194816 \\
11 &              &  4.611975415 &  8.828614846  & 13.49097737 &  18.47187696 \\
12 &              &  4.611975414 &  8.828614610  & 13.49097514 &  18.47126590 \\
13 &              &  4.611975414 &  8.828614592  & 13.49097251 &  18.47125237 \\
14 &              &              &  8.828614588  & 13.49097236 &  18.47123265 \\
15 &              &              &  8.828614587  & 13.49097230 &  18.47123172 \\
16 &              &              &  8.828614587  & 13.49097230 &  18.47123116 \\
17 &              &              &  8.828614587  &             &  18.47123111 \\

\hline
\end{tabular}
\end{center}
\end{table}

\begin{table}[tbp]
\caption{Eigenvalues for $V(x)=|x|^3$ and $g=-1$}
\label{tab:RRAO3_-1}
\begin{center}
\par
\begin{tabular}{|rD{.}{.}{10}D{.}{.}{9}D{.}{.}{9}D{.}{.}{8}D{.}{.}{8}|}
\hline $N$ &  \multicolumn{1}{c}{$W_0$}& \multicolumn{1}{c}{$W_1$}
&
\multicolumn{1}{c}{$W_2$}&\multicolumn{1}{c}{$W_3$}&\multicolumn{1}{c}{$W_4$}\vline\\
\hline

 2  & -0.2479044383 &  3.768307259  &              &             &               \\
 3  & -0.2753927098 &  3.767485564  & 8.274160995  &             &               \\
 4  & -0.2766897018 &  3.761603920  & 8.203629411  & 13.63427617 &               \\
 5  & -0.2768092707 &  3.760073668  & 8.152399613  & 13.08211481 &  20.30455979  \\
 6  & -0.2768240831 &  3.760009674  & 8.143450977  & 12.94688817 &  18.30513940  \\
 7  & -0.2768249160 &  3.759987489  & 8.143084282  & 12.89889957 &  18.13193414  \\
 8  & -0.2768249578 &  3.759987480  & 8.142733921  & 12.89731236 &  17.94860467  \\
 9  & -0.2768249641 &  3.759987026  & 8.142733513  & 12.89495715 &  17.94564705  \\
10  & -0.2768249646 &  3.759987019  & 8.142725848  & 12.89495536 &  17.93443115  \\
11  & -0.2768249646 &  3.759987013  & 8.142725786  & 12.89488104 &  17.93443096  \\
12  &               &  3.759987013  & 8.142725634  & 12.89488068 &  17.93396502  \\
13  &               &               & 8.142725629  & 12.89487879 &  17.93396223  \\
14  &               &               & 8.142725626  & 12.89487875 &  17.93394686  \\
15  &               &               & 8.142725626  & 12.89487871 &  17.93394655  \\
16  &               &               &              & 12.89487870 &  17.93394610  \\
17  &               &               &              & 12.89487870 &  17.93394608  \\

\hline
\end{tabular}
\end{center}
\end{table}

\end{document}